# ALLOCATING ACCESS TO QUANTUM COMPUTING: A LEGAL-ETHICAL FRAMEWORK

**Benedict Lane (TUD), Anushka Mittal (UvA), Ariana Torres-Knoop (SURF)**

KEYWORDS: LEGAL-ETHICAL FRAMEWORK, QUANTUM COMPUTING, VALUE CONFLICT, STAKEHOLDER ENGAGEMENT, ACCESS POLICIES

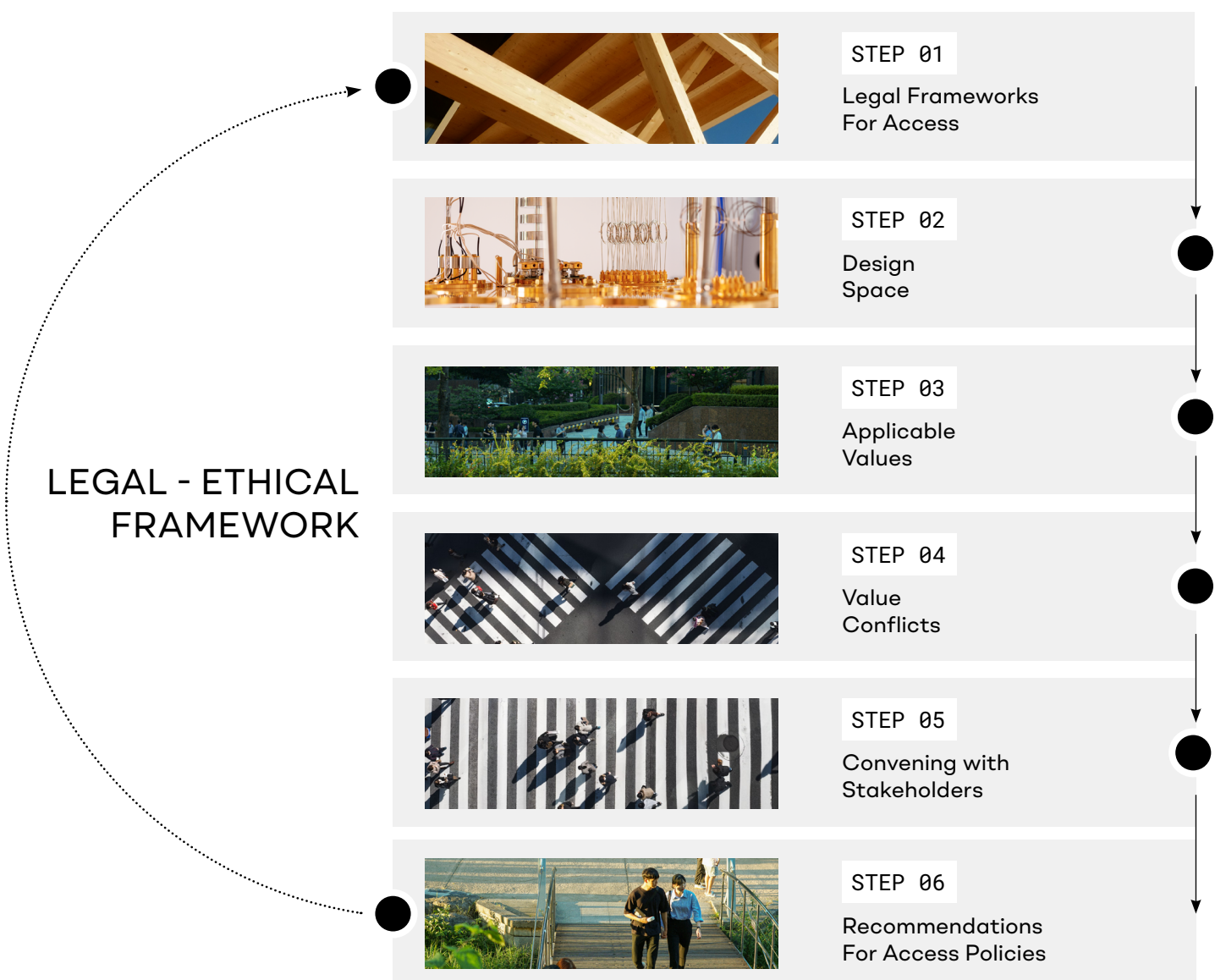

LEGAL - ETHICAL FRAMEWORK

**STEP 01**
Legal Frameworks For Access

**STEP 02**
Design Space

**STEP 03**
Applicable Values

**STEP 04**
Value Conflicts

**STEP 05**
Convening with Stakeholders

**STEP 06**
Recommendations For Access Policies

Quantum Delta NL

# 01. Introduction

## THE CONTEXT: THE RISE OF QUANTUM COMPUTING

Over the past decades, computing capacity has become not merely a tool but a foundational driver of scientific discovery and technological innovation. It is deeply embedded in a wide array of contemporary practices, and our dependence on it will only grow. Computing capacity forms part of the core infrastructure of modern life: drug discovery, climate modelling, factory optimisation, and advances in the automotive and aerospace industry – all rely on computing capacity to model, simulate and analyse data to deliver ground breaking scientific outcomes and advancing technologies. Its importance has grown even further as we enter the "fourth paradigm" of data-driven science, where discoveries increasingly emerge directly from experimental and simulation data rather than from progressively more complex theoretical models. [1]

National computing facilities provide universities, research centres, governments and industry with access to advanced computational and data-intensive methods, expert support, and cutting-edge infrastructure. They enable high-impact research by providing large-scale computing capacity and resources. In recent years, this has meant not only increasing the amount of capacity available but also diversifying the types of computing. Computing infrastructures are becoming more heterogeneous to accommodate the increasing demands of specialized workflows. Given that access to computing capacity is critical for innovation, economic growth and scientific progress, a coherent public policy agenda around this provision is a normative imperative for modern infrastructure planning. Within this evolving landscape, quantum computing is emerging as one of the most significant—and potentially disruptive—additions to the high-performance computing (HPC) ecosystem.

Quantum computing has sparked global interest for its potential to deliver exponential power for a limited but important set of applications. Until some years ago, quantum computers were largely confined to research laboratories and highly specialised expertise. In the last decade, rapid advances have brought them closer to integration to classical systems making them accessible through similar channels to those used in current computing paradigms – though they remain scarce and costly. While it is different from classical computing, it is increasingly integrated with it. This makes quantum computing follow the access channels of past computing paradigms.

Ensuring fair, strategic, and effective access to quantum computing within public and national infrastructures will be essential for sustaining competitive advantages in research, education, and innovation. As quantum and classical systems increasingly converge, the policy challenge will be to design allocation mechanisms that balance excellence, equity, and long-term societal benefit. Along with the increased availability and capabilities of quantum computers comes the core question: how can access to quantum computing be allocated in a responsible way?

## THE ISSUE: ALLOCATING ACCESS TO QUANTUM COMPUTING

There are many ways in which resources can be made available to users. This could be done by market mechanisms, state provisioning or collective sharing. These different channels impact the distribution and allocation of a resource as well as the claims over it. Once these questions are identified, the issue of allocating access to quantum computing becomes similar to that of distributing any other scarce resource: there is a fixed amount of some good, and it needs to be divided up among different actors, all of whom can have a claim to some portion of it. A prominent understanding of fairness in situations like this tells us that fairness requires us to satisfy claims in proportion with their strength. [2] In order to come up with a fair division, we therefore need to figure out who has a legitimate claim and how strong their claim is. For a good like quantum computing, we identify that legitimate claims and their strengths can be determined in two ways. The first is legal: who has a claim to a public good, and how strong their claim is to it, is partly determined by the law governing how that good is to be distributed and on what basis. When we talk about the law, we mean the relevant

AUGUST _ 2025

legislation itself, but also legally binding contracts governing access. The second is ethical. The law creates a "design space" for the possessor of a good to exercise some discretion in how they distribute it. Within the design space allowed for by the law and the limits imposed by the nature of the good itself, organisations and policymakers are faced with the task of coming up with a defensible way to assess the legitimacy and the strength of different claims, as well as a justifiable procedure for translating these claims into access. To do this, they must reflect on their own values, and identify the ethical considerations – the duties, values, and virtues – that bear on their choice. Both factors – legal and ethical – must be considered when allocating access to any public resource.

As various organisations think of providing access to quantum computing, their choices about allocation of access to quantum computing will be captured by the access policies and application procedures they finally adopt, as has been seen for other computational systems. The design of such policies and procedures would consider the context (legal, ethical, political, economic) in which the quantum system and computing infrastructure will be embedded.

What will be finally contained in an infrastructure's policies and procedures would be a distillation of many other laws, such as export controls on components and devices, national scientific agendas, and policies which prioritise certain applications, vendor-specific terms of services, and institutional terms of procurement and hosting, among others. The complex extant realities of establishing access policies at the intersection of multiple legal and regulatory frameworks is illustrated in Box 1.

<div style="border:1px solid blue">

## BOX 01 ACCESS TO HPC

As an example, let us consider the allocation of computing time to Snellius, the National Supercomputer in the Netherlands hosted and operated by SURF [3] (see Figure 1). Snellius is available for Dutch universities, research institutes and other organizations affiliated to the Dutch Research Council (NWO). Both, NWO Science (domain 'Exacte en Natuurwetenschappen') [4] and SURF are responsible for providing access rights (computing time) [5 & 6].

Access can be granted via:

1.  Grant applications: Large grants are evaluated by NWO and small grants can be directly requested to SURF.

2.  Direct contracts with the institutions

3.  European programs: e.g. SURF is the host of the European National Competence Centre (NCC) for HPC in the Netherlands (NCC Netherlands [7]).

4.  Co-investment in the infrastructure [8].

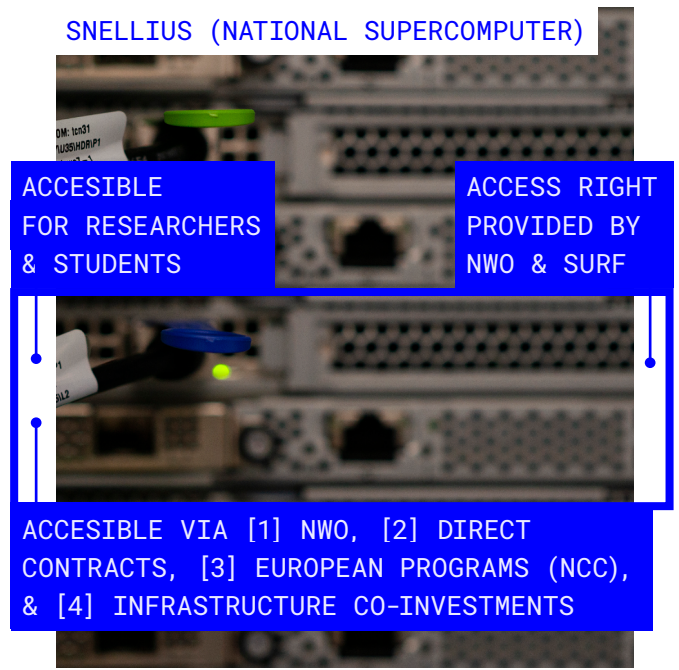

SNELLIUS (NATIONAL SUPERCOMPUTER)

ACCESIBLE FOR RESEARCHERS & STUDENTS

ACCESS RIGHT PROVIDED BY NWO & SURF

ACCESIBLE VIA [1] NWO, [2] DIRECT CONTRACTS, [3] EUROPEAN PROGRAMS (NCC), & [4] INFRASTRUCTURE CO-INVESTMENTS

FIGURE 1: The allocation of computing time of The National Supercomputer in the Netherlands, 'Snellius'.

</div>

## THE AIM OF THIS REPORT

This report introduces a general legal-ethical framework that providers of access to quantum computing can apply to develop robust access policies tailored to their specific context. We demonstrate the applicability of this general legal-ethical framework in the specific context of a small, 16-qubit quantum computer (QC) that will be hosted by SURF (the Dutch IT cooperative for research and education), integrated with Snellius (the Dutch national supercomputer), and operated jointly as part of the EuroSSQ-HPC consortium, procured in partnership with the EU-wide EuroHPC Joint Undertaking (JU).

Our application of the framework makes many simplifying assumptions and must be seen for what it is: a preliminary exploration of the legal and ethical issues bearing on the allocation of access to quantum computing and how these issues hang together. Our general assumptions include the (i) provision of publicly funded quantum computing (ii) to be placed in hosting organisations for scientific and research purposes.

There may be other distinctive ways in which quantum computing is provided. In view of the developments in this space, there is no pretence that these issues have been definitively dealt with, even in the limited context in which we consider them. Rather, our hope is that the way we approach these issues, in the narrow context in which we consider them, can be emulated and further refined by other organisations facing similar challenges in their own contexts. This has the potential to improve collective understanding of the structure of the challenges involved in allocating access to quantum computing and to identify context-transcendent best practices from which we will all benefit in the future.

We begin by laying down a step-by-step framework for thinking about defining access priorities and allocating access to quantum computing. This general framework is then applied to the case at hand, namely, the establishment of a 16-qubit quantum computer at SURF. Our application of the framework to this case is informed by the findings from a stakeholder workshop specifically conducted to address the issue of access. We follow this case study with some reflections, to highlight the limitations of the framework as well its potential to ignite an agenda for future research.

# 02.The legal-ethical framework

The following legal-ethical framework is intended as a device to facilitate providers of access to quantum computing to think through the issue of allocating access in a responsible and context-specific way. By separating out the relevant issues in a step-by-step process, the hope is that the multidimensional problem of designing an access policy for quantum computing can be tackled systematically (see Figure 2). The process ought not to be regarded as set in stone, but as a proposal, open to future modification and refinement in light of any limitations we discover by putting it into practice.

Our framework presupposes detailed knowledge of the technology to which access is to be distributed; we are imagining this framework being enacted after a procurement process has taken place, and therefore we are assuming that the hosting organisation will possess detailed understanding of the technology and its limits. This is important, because how technology ought to be used and allocated depends on how it can be used and allocated, and this is partly determined by the nature of the technology itself. For the purposes of designing an access policy, full and detailed knowledge of the micro-physical inner workings of a quantum computer is not necessary; however, knowledge of the functional capabilities of the computer is necessary, and ultimately this can only be derived from an understanding of the physics underlying the functionality of the machine. The policymaker need not possess this knowledge first-hand, but their functional understanding of the machine must be traceable back to such knowledge.

Furthermore, we assume a degree of knowledge of the potential for future developments of the technology; often expansion or upgrades are built into the hosting agreement. This is also important, because known future developments can affect the temporal evolution of priorities for allocating access. In this way, we assume both a certain "Technology Readiness Level" (TRL) but also what might be called a certain "Ethical Readiness Level" (ERL): that the technology is at an appropriate stage of technological determinacy for meaningful ethical questions to be asked about it. [9]

## LEGAL - ETHICAL FRAMEWORK

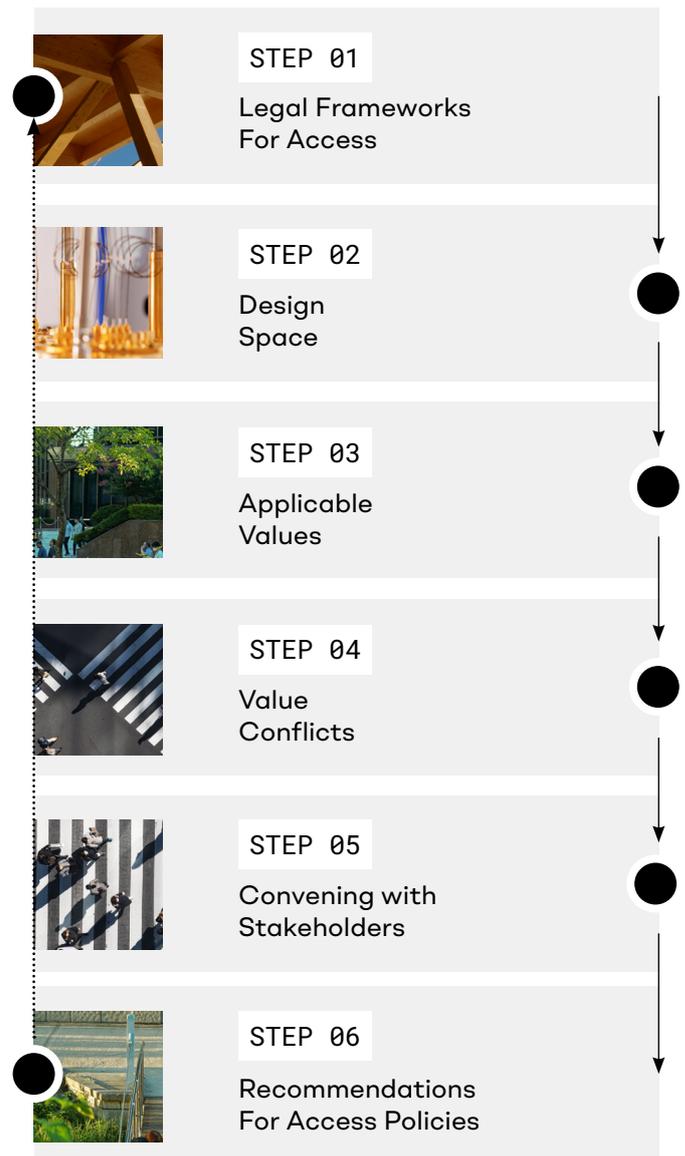

STEP 01

Legal Frameworks
For Access

STEP 02

Design
Space

STEP 03

Applicable
Values

STEP 04

Value
Conflicts

STEP 05

Convening with
Stakeholders

STEP 06

Recommendations
For Access Policies

FIGURE 2: The legal-ethical framework as a device to think through the issue of allocating access to quantum computers in a responsible and context-specific way.

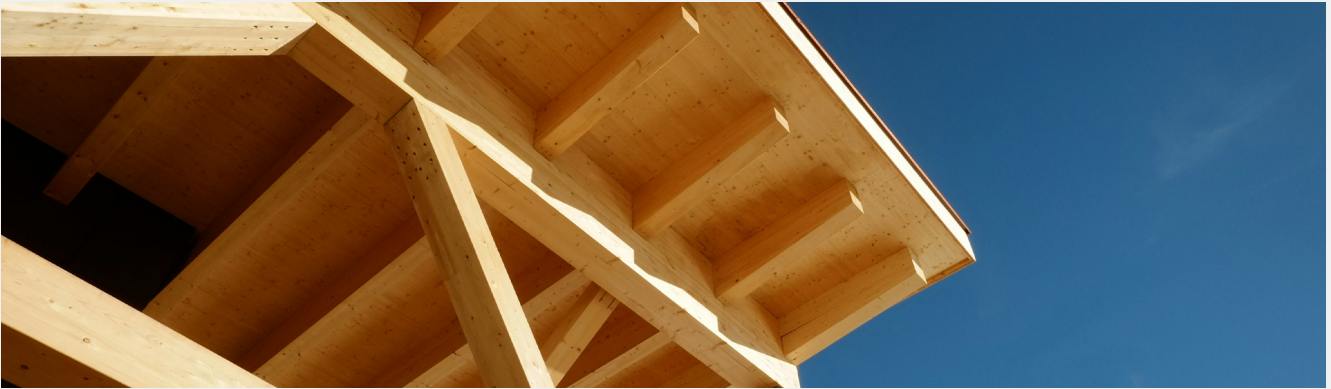

STEP 01

## LEGAL FRAMEWORKS FOR ACCESS

Due to their strategic and scientific significance, all quantum computers can be expected to be subject to a variety of legal and regulatory frameworks for access, directly or indirectly (see Figure 3). It can be subject to general legal frameworks, which might apply to a particular technology type (e.g., high-performance computing), and then specific ones, as developed for quantum computing. These frameworks can include contractual vendor restrictions, the policies and procedures of the providers (e.g. cloud and internet providers), the hosting organisation in case of publicly available quantum computers, as well as the national and supra-national legal and policy remit(s) under which the hosting organisation, the users, and the computing fall.

The first step for considering the issue of allocating access to a given quantum computer is for the applicable entity to look outwards to identify the frameworks of use or access that already apply to it, in terms of exploring various domains of laws.

VARIOUS DOMAINS OF LAW

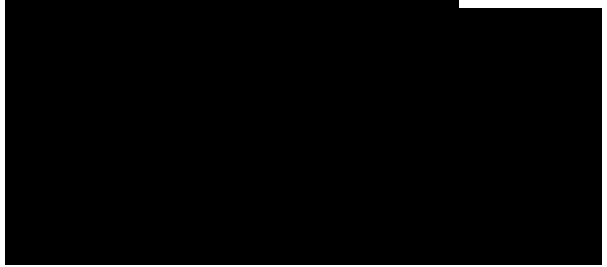

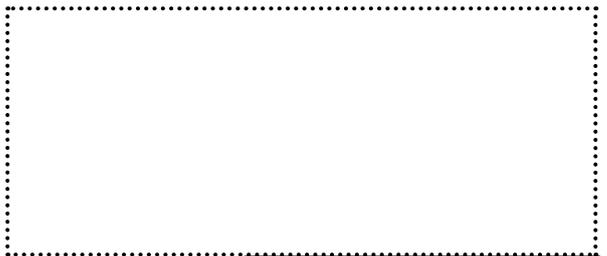

FIGURE 3: All quantum computers can be expected to be subject to a variety of legal and regulatory frameworks for access, directly or indirectly.

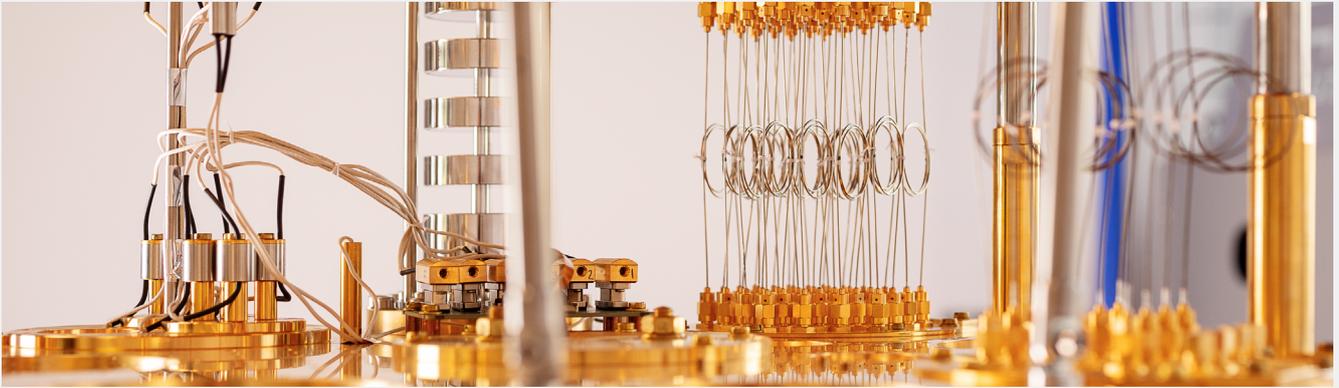

STEP 02

## DESIGN SPACE

The next step is to identify the degrees of freedom made available to the hosting organisation by the relevant legal and regulatory frameworks (and permitted by the nature of the technology and its readiness level).

This option space allows it to make credible decisions about access. We refer to these degrees of freedom as the "design space" available to a hosting organisation (see Figure 4), with the detailed and substantive elements of its access policy left to its discretion.

VARIOUS DOMAINS OF LAW

LEGAL FRAMEWORKS FOR ACCESS

'DESIGN SPACE'

REGULATORY FRAMEWORKS FOR ACCESS

FIGURE 4: The degrees of freedom made available to the hosting organisation is referred to as the 'design space'.

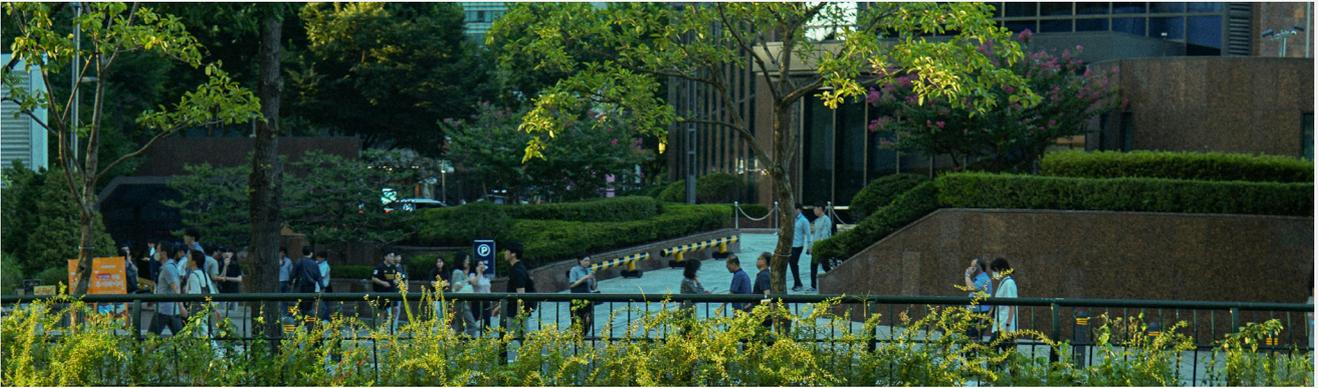

STEP 03

## APPLICABLE VALUES

Once a hosting organisation has a clear understanding of the design space available to it, the next step is to identify a principled basis for making choices about access within that design space. In this step, the hosting organisation looks both inward and outward to make explicit its own understanding of its obligations and responsibilities when it comes to allocating access to quantum computing. The hosting organisation will ask itself questions such as: what values define our mission? Whose interests are we responsible for championing? What consequences and outcomes fall under our remit? What norms and virtues define us as an organisation? It will also ask itself: what values are imposed on us from the outside? Which and whose external policies, priorities, and projects are we obligated to accommodate? Together, the answers to these questions constitute the hosting organisation's applicable values (see Figure 5).

An organisation's applicable values are not set in stone and can evolve when new challenges arise or when situations change – as indeed they are likely to do in the context of quantum computing. However, what differentiates a reasonable evolution of an organisation's applicable values from mere unprincipled action is a good-faith effort to articulate such values ahead of time. A key concept here is that of "reflexivity", which has been defined as "holding a mirror up to one's own activities, commitments and assumptions, being aware of the limits of knowledge and being mindful that a particular framing of an issue may not be universally held." [10]

HOSTING ORGANISATION

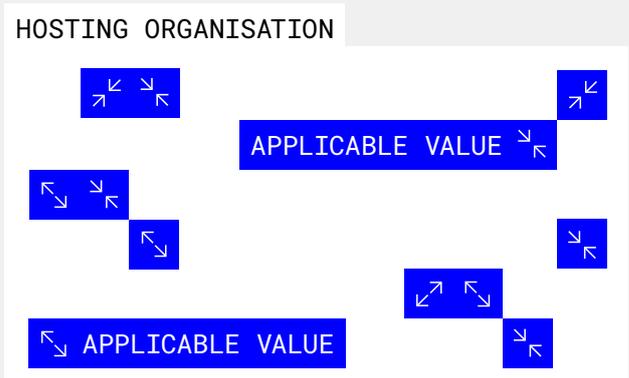

FIGURE 5: The hosting organisations' understanding of its obligations and responsibilities when it comes allocating access to quantum computing.

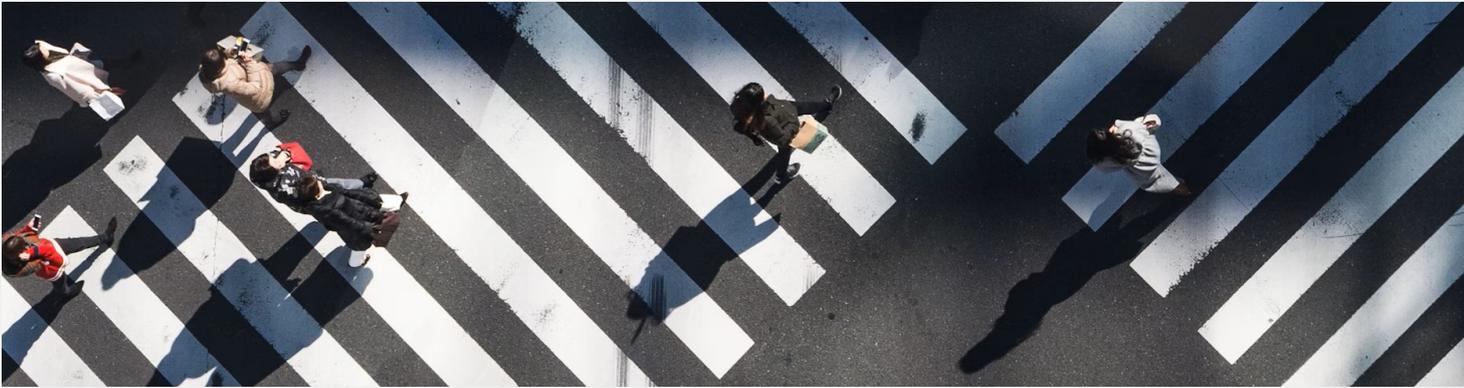

STEP 04

## VALUE CONFLICTS

With its applicable values explicitly articulated, the next step is for a hosting organisation to attempt to interpret and apply these values in the specific context of allocating access to quantum computing. For example, who has a legitimate claim to access the quantum computer through the host organisation, and for what purpose? Who has a legitimate stake in how access to the computer is allocated? How can access to the computer be allocated in a way that promotes the values that define the mission of the hosting organisation?

In the process of answering these questions, the hosting organisation will inevitably identify potential value conflicts, the choices open to the hosting organisation for allocating access about which their applicable values give incompatible or conflicting advice [11] (see Figure 6).

An organisation's values will further interact with other, more general ethical issues surrounding the technology to be distributed.

In the case of quantum technologies and quantum computing, these include:

- General issues about security and privacy: on the one hand, quantum computing may introduce a lack of transparency and accountability as a result of "blind computing"; on the other, QCs of sufficient scale and power can be a means for unethical uses of encryption breaking algorithms. These concerns around security and privacy are mutually reinforcing.

- Issues of distributive justice and the "quantum divide": quantum computing has the potential to exacerbate existing technical and educational inequalities and injustices through the localised and insular development of quantum technology. [12]

- Moral and prudential risks arising from the opacity of the future of quantum technology, including QCs: because quantum technologies are at such an early stage in their development, it is impossible to foresee the direction in which they will ultimately develop, and impossible to predict their eventual impact on society. [13]

These dilemmas, situated within the broader ethical landscape of quantum technologies, can help set the agenda for the host organisation's thinking about its access policies: they are the problems that the host organisation must solve in the process of developing its access policy.

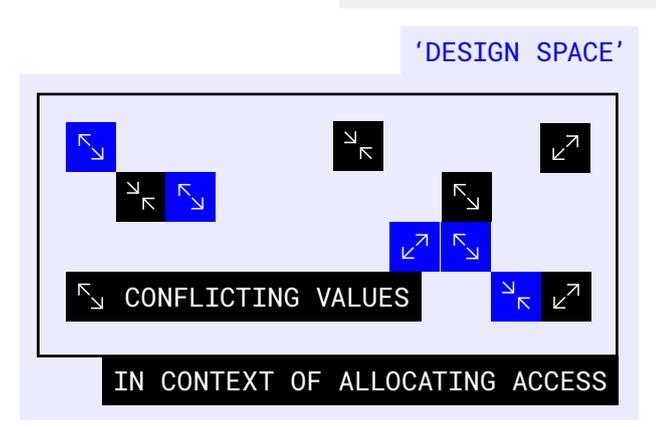

FIGURE 6: Conflicting values are identified by applying the applicable values of the hosting organisation in the context of allocating access.

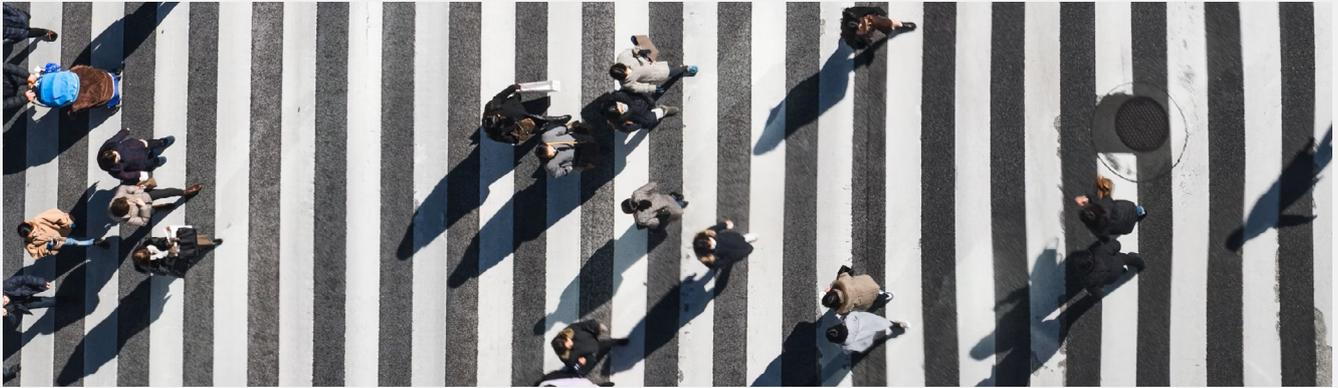

STEP 05

## CONVENING WITH STAKEHOLDERS

The upshot of steps 1-4 is a set of practical dilemmas, which will be unique for every hosting organisation. These dilemmas are then presented to the hosting organisation's stakeholders identified in step 4. Stakeholders are given a platform to interpret and apply the host organisation's applicable values to the issue of allocating access to quantum computing in their own way, drawing on their own individual understandings and perspectives (see Figure 7). In doing so, the identified value conflicts may find resolution: the choices over which the identified value conflicts arose may be reframed in such a way that the perceived conflict is eliminated; a clear ranking of values may emerge such that a particular choice emerges as most preferable; or an acceptable compromise between values may be identified such that each is satisfied to a sufficient degree; and so on. [14]

Convening with stakeholders serves two important functions. First, it plays a knowledge- and understanding-facilitating role: consultation can enable a host organisation to achieve insights that may have been invisible or inaccessible to it as a result of inflexible institutional habits or groupthink.

Second, it plays a normative, legitimating role: not only are the resolutions to the identified value conflicts better informed as the result of stakeholder engagement, but they are also better justified. Subjection to stakeholder scrutiny grants any resulting use-policy the badge of democratic legitimacy and thus contributes towards making that policy acceptable to prospective users. However, it must be emphasised that users and other indirect stakeholders are contributing advice; while they can both inform and legitimate a use-policy, they do not directly decide it. The power to decide lies with the host organisation.

STAKEHOLDER PLATFORM

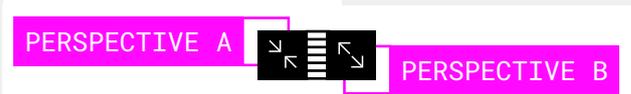

FIGURE 7: On the stakeholders platform the hosts conflicting values are interpreted and applied to quantum access allocation.

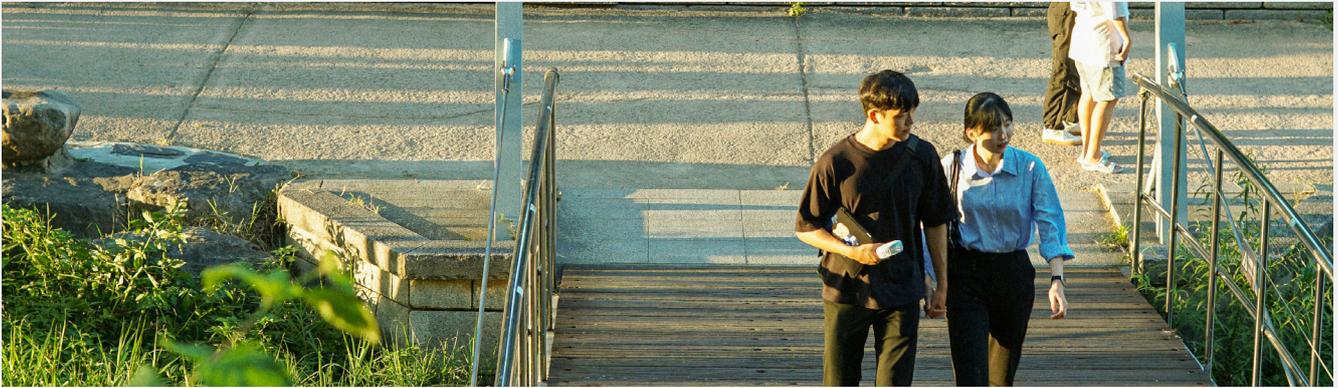

STEP 06

## RECOMMENDATIONS

Lastly, the solutions identified through the host organisation's engagement with stakeholders are translated into recommendations for an access policy (see figure 8). The value conflicts that fail to be resolved through engagement with stakeholders are fed back into step 3: the hosting organisation reconsiders its applicable values in light of the identified conflicts, and steps 4-6 are iterated.

HOST + STAKEHOLDER ENGAGEMENT

RESOLVED VALUE CONFLICT 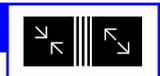

RESOLVED VALUE CONFLICT 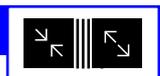

RESOLVED VALUE CONFLICT 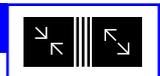

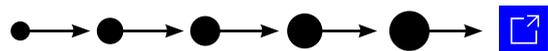

>> ACCESS POLICY RECOMMENDATIONS

FIGURE 8: Engagement outcomes are translated into recommendations for quantum computing access policies.

# 03.Case study: SURF

In this section, we apply the above legal-ethical framework to a specific case study: the 16-qubit quantum computer set to be hosted by SURF in 2026. Our case study makes a number of simplifying assumptions – an important example of which is addressed in Box 2. To arrive at a satisfactory access policy, these complicating factors would have to be reintroduced and accommodated. However, even our simplified test case turned up interesting results, and showed promise for the workability of the proposed legal-ethical framework.

## BOX 02 EUROSSQ-HPC CONSORTIUM PARTNERS

One simplifying assumption made in applying the developed legal-ethical framework to the SURF case was to consider only the values applicable to SURF. In reality, SURF is hosting its quantum computer as part of a consortium of Dutch, Belgian, and French institutional partners (see Figure 9). These include [1] the Netherlands eScience Center [15], [2] Nikhef (the Dutch national institute for particle physics) [16], [3] GENCI (the French counterpart to SURF) [17], and [4] universities in the Netherlands and Belgium (University of Leiden [18], Technical University of Delft [19], University of Antwerp [20]). While some of SURF's consortium partners might prioritise researchers from a variety of different fields, others might be more interested in attracting industry or talent development and education or quantum specific research.

There are many ways to undertake this exercise of determining values, ethical priorities and access shares which can essentially be grouped into (i) collectivising values and then determining access, or (ii) dividing up the access shares and then allocating each portion according to applicable values and mandates. Acknowledging the different sets of organisational values and applicable mandates- and identifying the potential conflicts between them – is fundamental to ensuring that the resulting access policy is to the benefit of all partners.

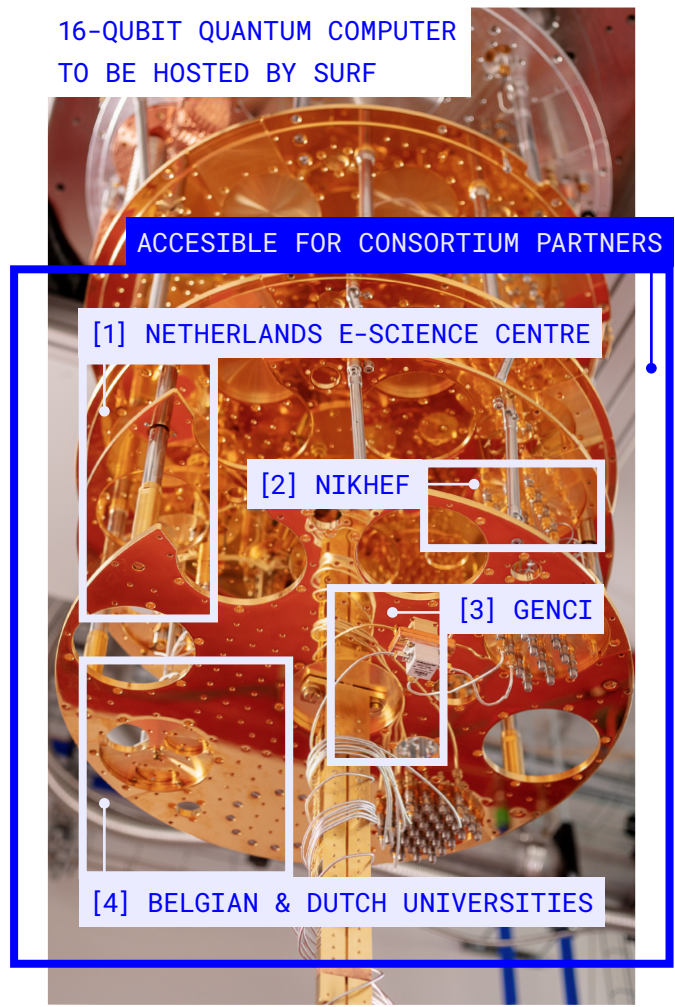

16-QUBIT QUANTUM COMPUTER TO BE HOSTED BY SURF

ACCESIBLE FOR CONSORTIUM PARTNERS

[1] NETHERLANDS E-SCIENCE CENTRE

[2] NIKHEF

[3] GENCI

[4] BELGIAN & DUTCH UNIVERSITIES

FIGURE 9: SURF is hosting its quantum computer as part of a consortium of Dutch, Belgian, and French institutional partners.

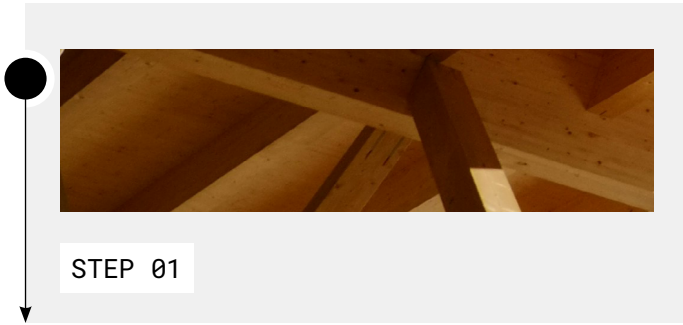

STEP 01

## SURF'S LEGAL FRAMEWORK FOR ACCES

The QC to be hosted by SURF is subject to the European High Performance Computing Joint Undertaking Regulation, 2021 (JU Regulation). It is a EuroHPC computer which has been procured by the consortium and the Joint Undertaking (JU). This law provides the possibilities, degrees of freedom, and precedents for how such advanced computing capacities have been allocated and accessed in Europe. Effectively, the EuroHPC ecosystem can be understood as an EU collective for EU's computing needs which is conditionally open for access to defined participants.

To explain its functioning in brief, the JU is organized as a public-private partnership. The JU pools monetary contributions from the European Union (45%), the member states [20] (45%) and private players (10%). It procures and operates various types of advanced computing systems and has an internal bureaucratic structure which manages the procurement of computers, the allocation of capacities, the prices charged and any liability issues. The procurement of the systems is based on the responses to open calls and the satisfaction of criteria like technical requirements, availability of power, electricity, network, etc., of interested "Hosting Entities" (in this case SURF).

The computers are owned wholly by the JU or partly with participating states and private parties. They are operated and managed by the Hosting Entity. For this, the Hosting Entity and the JU enter into a Hosting Agreement. The Hosting Entity further contributes to the costs of acquisition i.e. the remaining operational and maintenance costs. The distribution of these costs is important because it determines the shares in the computing capacity. The sequential order of steps for a quantum computer, as outlined in the JU Regulation, are provided in Figure 10 on page 15.

Under the JU Regulation, quantum computers are wholly owned by the JU and can only be hosted by participating states and their organisations (Article 12). For the quantum computer to be hosted at SURF, 50% of the share of computing capacity is held by the JU and the remaining is divided according to the share determined amongst the consortium partners (Recital 43, Article 17(4)). The JU's share of computing is distributed according to its Access Policy. This capacity is allocated in specific ways. The most relevant parameters have been identified below which guide the possibilities for SURF:

- Uses and Users – the Union share can be allocated for civilian uses, including dual use cybersecurity applications, to public and private entities within the EU. It is provided for free to conduct research and disseminate results based on open science and open access principles. The private entities are the ones supported by EU funds through Horizon Europe or Digital Europe grants while small and medium sized enterprises (SMEs) can get access regardless of their source of funds (Article 16, 17).

- External Access – the Union share can be allocated externally i.e. outside Europe under conditions of international cooperation and reciprocity (Recital 31) or for the benefit of the Union (Article 16(5)).

- Commercial Access – the Union share can be allocated commercially, capped to a maximum of 20% of its share (Article 18). This is paid access which circumvents the queue for peer review.

[20] The funding states are called Participating States.

AUGUST _ 2025

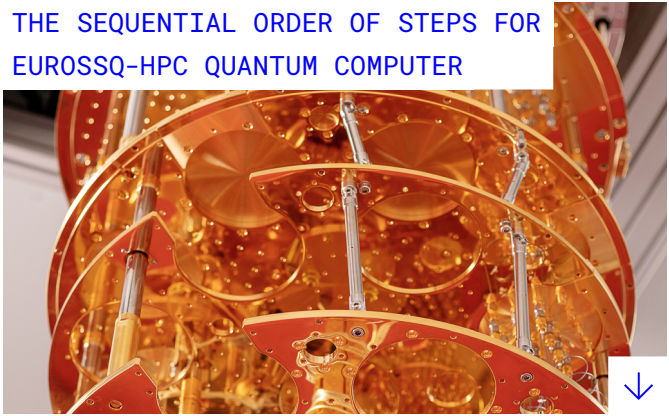

## THE SEQUENTIAL ORDER OF STEPS FOR EUROSSQ-HPC QUANTUM COMPUTER

### OWNERSHIP

The JU pools monetary contributions from the European Union (45%), the member states[1] (45%) and private players (10%).
[100% JU]

### HOSTING

The procurement of the systems is based on the responses to open calls and the satisfaction of criteria.
[100% EUROSSQ-HPC]

### ACQUISITION

The distribution of these costs is important because it determines the shares in the computing capacity.
[50% JU, 50% EUROSSQ-HPC]

### ACCESS

The share of computing capacity is held for 50% by the JU and 50% by the consortium partners.
[50% JU, 50% EUROSSQ-HPC]

FIGURE 10: The sequential order of steps for the QC to be hosted at SURF which is subject to the European High Performance Computing Joint Undertaking Regulation, 2021 (JU Regulation).

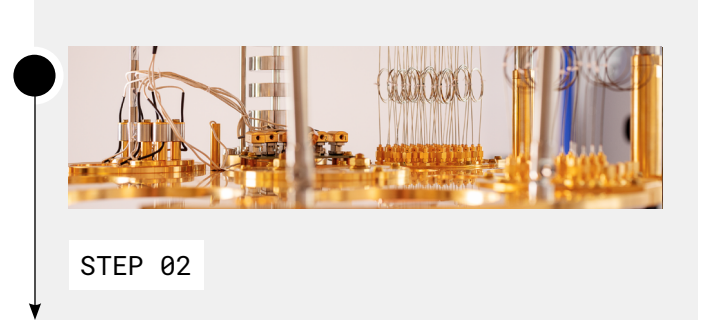

STEP 02

## SURF'S DESIGN SPACE

To identify all possible limits, we reviewed the JU Regulation, draft Hosting Agreement and JU Access Policy. In this case, these documents reinforce each other and lay down a prescription of what can be done with the computers the JU funds, including the QC to be hosted at SURF. As explained, the Union's share is decided by the JU Regulation in a formulaic way while the consortium partners can decide claims for their share (the remaining 50%). The design space exists in this remaining share. As SURF and the consortium partners are bound together to the JU (which is the owner, co-funder, regulator of QC), their share is necessarily inspired, guided or bound by the JU's prescriptions, however interpreted. [21]

This design space is flexible in some ways and inflexible in others. For instance, the consortium can decide on the possibility and allocation of computing time for commercial access. At the same time, it would not be advisable to grant external access to entities beyond what the JU decides. These aspects may further be guided by general good practices and even involve political and governmental actors. This space requires a discussion between consortium members. For our purposes, the law applicable to the Union's share (JU Regulation, Access Policy), the relationship of the consortium with the JU (JU Regulation) and SURF as the Hosting Entity (Hosting Agreement, SURF's values) provided a fertile ground to substantively discuss the available options within the design space available to SURF. There may be other kinds of documents such as End User Agreements, Proposal calls, international bilateral agreements which can be useful to understand this environment better. Depending on the applicability and availability of documents, other institutions can benefit from identification and review of key documents.

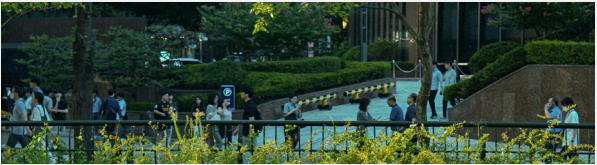

STEP 03

## SURF'S APPLICABLE VALUES

Conveniently, SURF has previously identified its organisational values with regard to the digital transformation of education and research. [22] Appropriately adapted, these values can steer SURF's thinking about its ethical obligations when it comes to allocating access to quantum computing as well. Of course, these values can be questioned, revised, and refined in light of new insights and challenges that arise in this context - they serve only as a starting point.

In this case, three values are identified. The first of these values is autonomy. Autonomy is a complex value, the dimensions of which include freedom of choice, privacy, and independence – for SURF, these apply to individual students, researchers, and teachers, but also to institutions themselves. The second is justice. Justice is no less multifaceted, encompassing "substantive" goods such as inclusivity, accessibility, equality of opportunity, equity of outcomes – again, at both the individual and institutional levels – as well as "procedural" values such as transparency, democracy, and trustworthiness. The third is humanity. Promoting the value of humanity involves engaging stakeholders respectfully, engendering the personal development of and meaningful contact between stakeholders, fostering individual flourishing and social cohesion more broadly, as well as a commitment to sustainability.

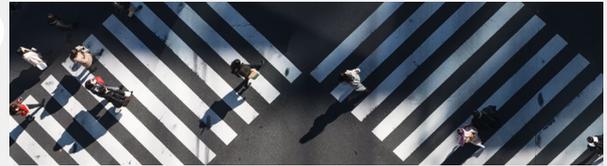

STEP 04

## POTENTIAL CONFLICTS BETWEEN SURF'S APPLICABLE VALUES

The interpretation and application of SURF's organisational values in the context of allocating access to quantum computing results in (at least) the following dilemmas:

- Intuitively, the values of justice and humanity tell in favour of open science and promoting international participation and collaboration. This may entail allocating a significant proportion of available access to the quantum computer to users outside of EuroSSQ-HPC consortium countries, or even outside of the EU. The value of autonomy, however, might favour allocations to promote independence, interpreted as the technological sovereignty and self-reliance of consortium countries and of the EU.

- Reasoning based on the value of autonomy might endorse allocations based solely on scientific merit. However, this may disproportionately benefit users with the most pre-existing knowledge, experience, and resources. In contrast, the value of justice emphasises equal opportunity, equitable outcomes, and inclusivity, and the value of humanity emphasises the development of new talent and knowledge.

- Promoting procedural values such as transparency and democracy may involve SURF collecting or even publicly disclosing certain information about who is using their quantum computer and how they are using it. However, this might be perceived as compromising the privacy and security of SURF's clients, especially those conducting sensitive or proprietary research, and therefore in tension with their autonomy.

- The value of humanity may be open to two mutually unrealisable interpretations simultaneously. Research with clear, immediate social benefits – e.g., the discovery of new drugs and materials – might divert resources away from more exploratory research with the potential to drive beneficial innovation in the long run. This creates a dilemma when interpreting SURF's commitment to sustainability and human flourishing in this context.

The issue of value conflict is exacerbated by the interaction of the specific ethical challenges faced by SURF and its consortium partners with the more general ethical implications of quantum computing mentioned above: security and privacy issues; distributive justice and the "quantum divide"; and the uncertain future development and impact of quantum technologies.

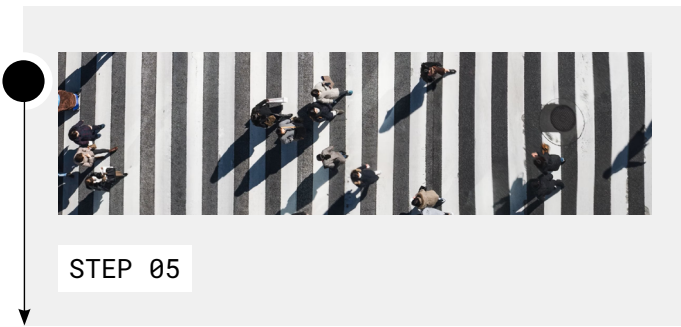

STEP 05

## CONVENING WITH SURF'S STAKEHOLDERS

The identified aspects of the regulatory regime and ethical values provided the template to engage with relevant stakeholders. We conducted a workshop at SURF on February 6, 2025. Participants in the workshop came from SURF itself, research institutions in the Netherlands and Belgium, the Dutch government (the Rijksinspectie Digitale Infrastructuur), and the Centre for Quantum and Society within the Dutch quantum ecosystem QuantumDeltaNL. Participants were presented with dilemmas identified in Step 2 and given the opportunity to exchange their views about the allocation of the quantum computer to be hosted by SURF along the dimensions outlined in Step 1 (external, commercial, and use-cases). We were mindful of the Union's choices and SURF's values and invited participants to bridge the space between a real-life policy and a hypothetical mental exercise.

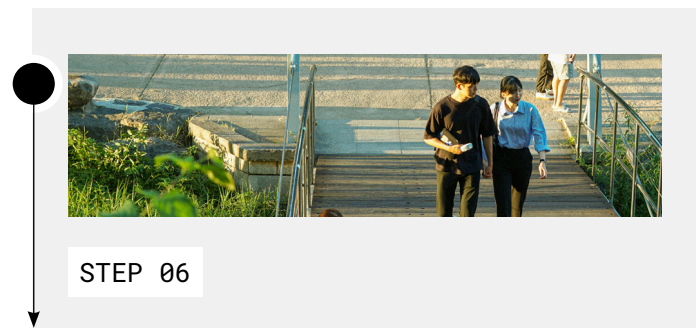

STEP 06

## RECOMMENDATIONS FOR AN ACCESS POLICY

The results of the Workshop are reported below, condensed into substantive themes for SURF and its consortium partners to potentially use as policy inputs as they draft an access policy for the QC. These thematic priorities for access arise given the contextual constraints applicable to the QC.

### Theme 1: The regional value of global and commercial partnerships

Participants recognised a tension between the value of regional autonomy and the value of open science. Issues of knowledge security, the fairness of benefits from EU funding going to non-EU users, and legal obligations of QC vendors, such as export controls, were also positioned as in tension with the goals of open science and promoting international participation and collaboration. However, several practical suggestions were made for reframing and resolving this tension. It was recognised by many participants that global cooperation under certain circumstances may be a way of building regional autonomy. Along these lines, the following suggestions were made:

- Allocating external access by issuing "themed calls" for proposals that address knowledge gaps and practical issues that matter to regional stakeholders. This would advance regional interests while building a global network and increasing the visibility of European quantum computing providers.

- Allocating external access based on open-source contributions - i.e., on the basis of the quality of transferable results that can be made accessible to regional users.

- Allocating external access with the goal of encouraging mutual talent development and iterative collaboration - i.e., on the basis of the benefits to regional institutions from building constructive, long-term relationships with external institutions.

- Raising the global profile of regional and European quantum by instituting a crediting system, whereby external users are obligated to credit SURF and its consortium partners for results arising from the QC access they are granted.

## Theme 2: The temporal evolution of both technology and value priorities in time

A theme that repeatedly arose was the evolution in time of priorities regarding access. Priorities were recognised as changing over time alongside changes in: (i) the technology itself, e.g., the number of available qubits, given that SURF's QC will be slated for expansion after two years; (ii) technical understanding and know-how, e.g., the discovery of new applications and better awareness of the limitations of the hardware; (iii) user demand and use patterns. Acknowledging this led to several original suggestions for reframing and resolving SURF's value dilemmas.

One aspect of the temporal evolution of priorities concerns how to best categorise different use cases as time goes on. Participants identified several different systems of "pigeonholes" for categorising use cases (see Figure 11). First, categorisation based on maturity-level. When categorised this way, uses are grouped according to their contribution to (1) talent and knowledge development, (2) proof of concept, or (3) market readiness. Second, categorisation based on domain. When categorised this way, uses are grouped according to the field of research to which they contribute, such as (1') education, (2') physics, chemistry, biology, and materials science, (3') optimisation, (4') quantum AI, and so on. Third, categorisation according to purpose. When categorised this way, uses are grouped according to their intended real-world function, for example (1'') military use, (2'') civilian use, or (3'') dual use.

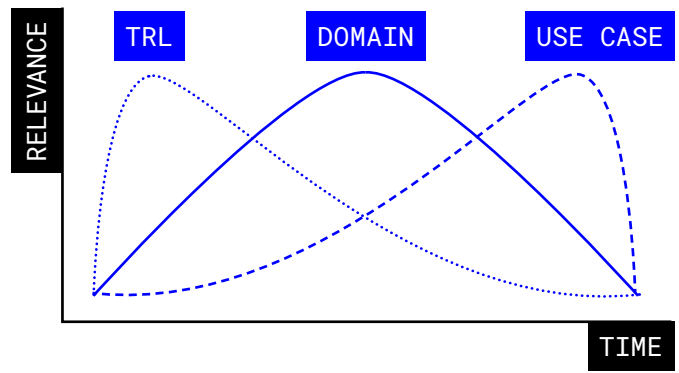

FIGURE 11: Several different systems of "pigeonholes" for categorising use cases were identified by the participants of the stakeholder platform workshop on 6th of February 2025.

Which of these systems of "pigeonholes" is going to be most useful for making choices about allocation is going to change with time. For example, in the early stages, maturity-level categories are going to be most salient because the goal is to advance quantum technologies to a higher stage of maturity. Once a strong talent- and knowledge-base is established, use-domains may gain momentum. Domain-based categories will then take over in importance, since access will have to be allocated between research in different use-domains, each vying for use of the machine for proof-of-concept of domain-specific applications. Once an array of domain-specific applications come to fruition, purpose-based categories will then take over in importance, since access will have to be allocated between users vying for use of the machine to employ domain-specific applications for different user-specific purposes.

Participants were presented with the dilemma between prioritising scientific excellence and prioritising the development of new talent. Given the evolution of priorities described above, one suggestion was for allocation to initially prioritise the goal of developing a strong talent- and knowledge-base, by (a) providing clients with education and hands-on experience of using basic quantum computing hardware, and (b) developing knowledge of the capabilities and limitations of the hardware and software itself, such as benchmarking and error mitigation. Once experience, knowledge, and technology develop, emphasis can shift to developing higher-impact applications in different use-domains based on scientific excellence. Thus, by prioritising talent development and scientific excellence sequentially, the dilemma of prioritising one over the other is partially resolved.

Another complementary suggestion for resolving this same dilemma was for SURF to emulate a "CERN model" [23] for allocating access: given that the maturity level of the relevant technology will increase over time, the most up-to-date resources can be devoted to high-quality, cutting-edge research, while obsolete "previous versions" of software and hardware can be made available for public use, thereby fostering new talent. This way, as both technology and understanding advance, access can be opened up to more exploratory, less cutting-edge uses. By recognising the temporal evolution of both the technology and value priorities, the value conflict between excellence and inclusivity can again be partially resolved.

**Theme 3: Paid use as a means-to-an-end**
When presented with the possibility of paid use of the QC to be hosted by SURF, participants discussed the allocation of access to commercial users in quite "Rawlsian" terms [24] (see Figure 12), in the sense that a strong emphasis was placed on using money gained by paid use to improve both the capabilities of the machine itself as well as the state of quantum regionally, for the benefit of less resourceful, non-commercial users. Commercial (i.e., paid-for) access was also perceived as a necessary return on investment for the EU and for consortium partners. Conversely, it was emphasised that the cost of (i) the overhead for providing commercial use and (ii) meeting the "service expectations" of a commercial relationship needs to be counterbalanced by benefits. i.e., paid use would be offered to the extent that, and no more than, it would benefit other potential users.

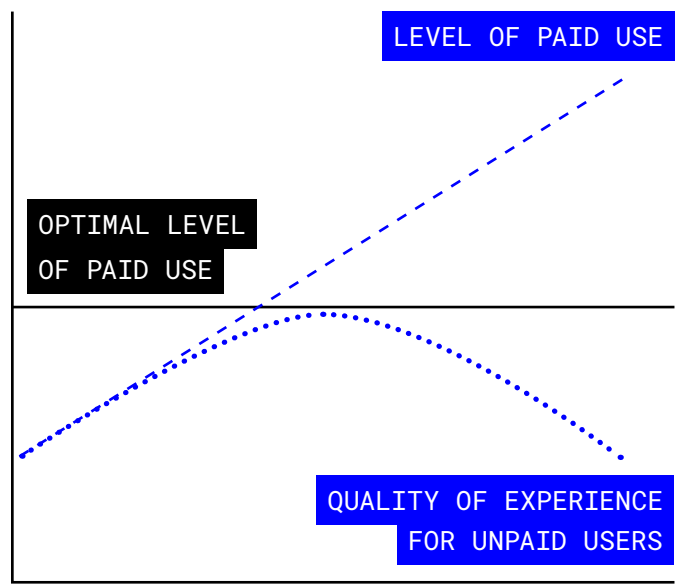

FIGURE 12: Rawlsian "difference principle" showing how the "worst off" user can be best off as a result of limited paid use

Because the initial computer will be so small, one suggestion was that initially commercial access be prioritised for non-profit-making private-sector R&D, before shifting to profit-making commercial use (in line with the temporal evolution of priorities mentioned above). There was also an emphasis that "profit" was not a dirty word, and that for-profit commercial use of QC benefitted regional actors through tax revenue, private-sector talent development, and driving innovation. However, opinions among participants varied, with others suggesting that SURF and its consortium partners ought to capitalise on the scarcity of QC to maximise financial return on their investment in the machine.

### Theme 4: Concerns about enforcement

Participants were presented with the dilemma between the procedural values of transparency, democracy, and accountability and the value of users' autonomy. A key set of concerns identified by participants had to do with enforcement of any eventual access policy. Participants identified conceptual, technical, and ethical issues:

- Conceptual issues had to do with defining and operationalizing key terms and categories on the basis of which use is allocated. For example, if initially SURF and its EuroSSQ-HPC partners chose to prioritise commercial access to R&D - as suggested above - how could R&D be defined in a consistent and operationalizable way that isn't subject to exploitation through semantic loopholes?

- Technical issues had to do with how to monitor and apply a particular use policy, given the way that the quantum computer will physically work. For example, by design, it is impossible to make observations of the execution of quantum algorithms without disrupting them. How can SURF make sure such a machine is being used appropriately?

- Ethical issues had to do with SURF's responsibility for the use to which the computer is put, as well as its duty to respect the autonomy of its users and to avoid acting paternalistically.

Each of these issues were exacerbated by a lack of existing discussions for monitoring or controlling the outcomes or purposes of use once access has already been allocated and granted to HPC infrastructures more generally.

This dilemma proved especially difficult to overcome. One suggestion was not to have users submit their code directly (this was deemed too paternalistic, violating autonomy) but to offer detailed and clear (terms that will themselves require consistent definition and operationalisation) plans for what they intend to do with the machine – this adds an additional "effort barrier" to misuse, a "soft" way for SURF to police use of the machine. This suggestion doesn't represent a satisfying resolution of the dilemma, but rather an attempt to cope with it. Another proposal was to have a technical assessment of proposals (conducted by an expert panel) to assess whether this was, at the very least, a good use of the machine's time, or if a high-performance classical computer or a classically simulated QC (e.g. Quantum Inspire) would be more suitable. In this way, the proposed review process would be constructive, helping to get the best out of a given proposal, rather than an exercise in micromanaging; enforcement is thereby carried out in a way that promotes, rather than violates, users' autonomy, representing a partial resolution of the dilemma.

# 04. Reflections

The case study results in some open questions which also have a bearing on thinking about access in a responsible way. For instance, is it realistic to expect any ex ante risk analysis to capture all of the possible undesirable uses to which the machine might be put? Can unaffiliated external researchers access the capacity provided by SURF? To what extent can SURF, as the Hosting Entity and gatekeeper of capacities, influence the future uses? Can a new regime be thought of for such advanced and powerful computing, for different stages of technical maturity?

## UNEXPLORED ISSUES

In this report, we explored only the tip of a vast iceberg, putting to one side the full details of the setting, country, region, laws and values that bear on the issue of allocating access to quantum computing. We discussed the availability of quantum computing in the Netherlands, which is subject to the considerations of a consortium, its institutions and the larger European context. Apart from the factual assumptions, we did not consider the following substantive issues:

### Impact of European research infrastructure on access

We did not consider the impact of an ecosystem of computing, across Europe. In the space of quantum computing, there are now 8 (6 [25]+ 2 [26]) centers, at different levels of progression, with different types of quantum computers and integration with other classical systems. These systems are embedded in larger e-science and research infrastructures in the EU. This includes distinct research networks, data infrastructures, cloud services and computing centers which have their own boundaries. These elements connect particular institutions and technical systems which finally determine the access for individual researchers. A federation of such systems is also envisaged in the future which was not covered by our work here. These complicating factors make it important to reiterate that allocating access to advanced computing resources is highly contextual.

### Technical and institutional integration with existing systems

Furthermore, the scope of the workshop in which we tested the proposed framework was restricted to discussing the quantum computer in isolation, focusing on its known characteristics. However, in reality, the QC will be integrated with the Dutch national supercomputer, Snellius. This introduces additional channels of access, via the Dutch Research Council, NWO. Thus, the question of allocation gets further complicated, potentially introducing a need to align the technological possibilities and value priorities of allocating access to Snellius and the QC.

## AGENDA FOR THE FUTURE

This report recognises the importance of addressing the problem of access to quantum computing and is intended to initiate a conversation. Further questions remain. For example:

- What recourse should there be for users who are overlooked by an access policy, and miss out on accessing quantum computing?

- Given the public money that goes into creating and sustaining quantum infrastructure, which institutions ought to be responsible for making decisions about access?

- Are there alternative channels for access to quantum computing that we have not considered but which need to be taken into account?

We hope the conversation we have initiated will turn to addressing these questions moving forward, incorporating the voices of as many stakeholders as possible.

The bridge between laws and values drawn here is one of the many ways to think about access to computing resources, grounded in political, economic and technical realities. These are the identified ingredients to turn a mental exercise into a policy one. They are necessary but not always sufficient for arriving at an acceptable access policy.

Lastly, it is important to emphasise that the legal-ethical framework developed and applied in this report is intended to be iterative, contextual and temporal. The issue of allocation and access requires continuous reconsideration as circumstances change. These circumstances range from the stage of technology, value priorities, organisational and geopolitical alliances to institutional or national contexts. Europe provides a mature framework to discuss these issues in the form of a research ecosystem, infrastructure, computational resources and policy priorities. Changes in these contextual factors can lead to distinct questions and considerations which need a wider and deeper discussion.

# 05. Acknowledgements

This paper and the workshop described are part of the activities of a working group composed by ECP, QDNL CQS, Rathenau Institute, SURF, TUD, TNO and the UvA. We thank the participants of the workshop for their active participation and input and our colleagues for fruitful discussions and reviewing.

Benedict Lane's contribution to this work was supported by the Dutch National Growth Fund (NGF), as part of the Quantum Delta NL programme.

Anushka Mittal's contribution to this work was supported by Action Line 4 as part of the Quantum Delta Netherlands programme.

Graphic design of this paper was created by ECP and Juwe van Vliet, with images of Bernhard (via Unsplash), Roji Iwata (via Unsplash), Jieun Kim (via Unsplash), Marieke de Lorijn, Snellius (the National Supercomputer), and David Trinks (via Unsplash).